# All semiconductor enhanced high-harmonic generation from a single nano-structure

Dominik Franz[1], Shatha Kaassamani[1], David Gauthier[1], Rana Nicolas[1], Maria Kholodtsova[1], Ludovic Douillard[2], Jean-Thomas Gomes[3], Laure Lavoute[3], Dmitry Gaponov[3], Nicolas Ducros[3], Sebastien Fevrier[3,4], Jens Biegert[5], Liping Shi[6], Milutin Kovacev[6], Willem Boutu[1] and Hamed Merdji[1*]

[1] *LIDYL, CEA, CNRS, Université Paris-Saclay, CEA Saclay 91191 Gif sur Yvette France*

[2] *SPEC, CEA, CNRS, Université Paris-Saclay, CEA Saclay 91191 Gif sur Yvette France*

[3] *Novae, ZA du Moulin Cheyroux, 87700 Aixe-sur-Vienne, France*

[4] *Univ. Limoges, CNRS, XLIM, UMR 7252, 87000 Limoges, France*

[5] *ICFO – The Institute of Photonic Sciences,*
*Mediterranean Technology Park, Av. Carl Friedrich Gauss 3, 08860 Castelldefels, Spain*

[6] *Leibniz Universität Hannover, Institut für Quantenoptik, Welfengarten 1, D-30167 Hannover, Germany*

[*]*correspondence to: hamed.merdji@cea.fr*

**The enhancement and control of non-linear phenomena at a nanometer scale has a wide range of applications in science and in industry. Among these phenomena, high-harmonic generation in solids is a recent focus of research to realize next generation petahertz optoelectronic devices or compact all solid state EUV sources. Here, we report on the realization of the first nanoscale high harmonic source. The strong field regime is reached by confining the electric field from a few nanojoules femtosecond laser in a single 3D semiconductor waveguide. We reveal a strong competition between enhancement of coherent harmonics and incoherent fluorescence favored by excitonic processes. However, far from the band edge, clear enhancement of the harmonic emission is reported with a robust sustainability offering a compact nanosource for applications. We illustrate the potential of our harmonic nano-device by performing a coherent diffractive imaging experiment. Ultra-compact UV/X-ray nanoprobes are foreseen to have other potential applications such as petahertz electronics, nano-tomography or nano-medicine.**

Ultrafast nano-photonics science is emerging thanks to the extraordinary progresses in nano-fabrication and ultrafast laser science. Boosting laser fields in nano-structured photonic devices to the strong field regime has the potential of creating nano-localized sources of energetic photons or particles, opening vast applications. Nowadays, optoelectronics is extending to the highly non-linear regime where electron currents can be optically controlled under strong fields in semiconductors. A recent impact of this capability is the emergence of high-harmonic generation (HHG) in crystals[1–5]. High harmonics are emitted when electrons undergo either intra-band or inter-band processes. These above bandgap phenomena occur efficiently in a crystal exit layer of sub-micrometer thickness[1] down to an atomically thin layer[6–10]. The strong electron currents from which HHG originates can be manipulated in space and time. This, in turn, allows to localize the HHG process in time at the single optical cycle scale[4] and in space at a nanometer scale, as presented in this work. This control can not only revolutionize attosecond science but also prepare a new generation of ultrafast visible to X-ray optoelectronic devices operating at petahertz frequencies.

The high intensity needed for HHG in solids requires large laser amplification systems which limits the integration of the source in advanced devices. However, nanoscale field enhancement is nowadays a technology used to stimulate non-linear phenomena[11,12]. Different meta-surfaces of metallic[13–15], dielectric and semiconductor[16–18] as well as metallic/dielectric and metallic/semiconductor hybrid structures[19–21] were successfully used to boost non-linear processes. In 2016, Han et al. showed the enhancement of high harmonics of an 800 nm femtosecond oscillator due to plasmonic resonances in hybrid sapphire-metal waveguides[22]. More recently, enhancement of high-harmonic generation induced by metallic nano-rods on a silicon thin film has been observed[23]. Even though plasmonic resonances ensure the highest enhancement of the local electric field, still, both works report major limitations. First, metallic nano-structures are easily damaged at high intensities; second, the enhancement volume is extremely small and, finally, transmission of high harmonics through the metal is very low, which limits the non-linear conversion efficiency. Another route has been proposed by Sivis et al. that paves the way for HHG enhancement in semiconductors[24]. However, enhancement in 2D grating structures was only demonstrated for light emitted around the crystal bandgap which makes it difficult to disentangle coherent harmonic versus incoherent luminescence enhancement. Liu et al. have recently reported on the enhancement of harmonics using an array of structures patterned on an all-dielectric 2D surface[25]. Here, we report advances on the enhancement of above band gap harmonics in an isolated 3D semiconductor waveguide. The nanoscale localization of the harmonic emission is reported for the first time, opening the path for a compact ultrafast extreme UV nano-probe.

**Results**

The principle is based on the confinement of the laser light in an integrated structure to reach the intensity threshold that allows for high-harmonic generation from a few nanojoules mid-infrared fiber laser. The effective laser intensity is raised up from the sub-TW/cm$^2$ to the TW/cm$^2$ regime inside a single zinc oxide (ZnO) nano-waveguide as illustrated in Fig. 1a. The design of the structure and the results from finite-difference time-domain (FDTD) simulations of the intensity enhancement are shown in Fig. 1b. As the beam propagates towards the tip of the ZnO nanocone, the fundamental intensity is enhanced by more than one order of magnitude in a sub-laser-wavelength volume. The simulation of the laser confinement at the output surface in the plane perpendicular to the propagation direction shows a regular centered mode (see Fig. 1c). Note that intensity enhancement close to the exit surface is mandatory for efficient above bandgap HHG because high harmonics cannot propagate more than few 100 nanometers in the bulk crystal due to their strong absorption[1].

The layout of the experimental setup is shown in Fig. 2 (see methods). Few nanojoules are coupled into the nanocone entrance to seed the harmonic emission. Using a UV/VIS spectrometer, we have measured emission spectra at different pump intensities. The results are displayed in Figs. 3a and 3b. At low pump intensity of 0.07 TW/cm$^2$ only H3 is generated from the bare crystal and emission from the defect states (450-550 nm band) is noticeable (blue curve, Fig. 3a). While this intensity is not sufficient to trigger above-band gap harmonic generation in the bare crystal, H5, H7 and H9 are efficiently generated when the laser is focused into the nanocone (red curve, Fig. 3a). The supplementary movie S1 shows the dynamic enhancement when the isolated truncated nanocone is scanned across the laser beam focus. In Fig. 3a, we notice a very strong enhancement of the luminescence generated from the excitonic band edge (385 nm) and the defect states spectral regions. When increasing the pump intensity to 0.17 TW/cm$^2$, we notice that H5, H7 and H9 and the luminescence are generated from the bare crystal (blue curve, Fig. 3b). When focusing into the nanocone, all harmonic orders and the luminescence are dramatically enhanced with respect to the bare crystal (red curve, Fig. 3b). We see that the luminescence from the band edge, which could barely be measured from the bare crystal, is strongly enhanced so that H5 is almost totally immersed in this incoherent signal. Moreover, Fig. 3c reports the non-perturbative intensity dependence of the luminescence which scales as $I^{3.1}$. This does not allow us to clearly identify the strong field origin of the harmonic emission such as reported by Sivis et al.[24].

In order to discriminate between coherent and incoherent enhancement, we concentrated our study on higher harmonic orders. H7 (300 nm) and H9 (233 nm) stand far from the ZnO excitonic band edge which allows for a clear observation of the coherent enhancement process. H7 and H9 are enhanced in the nanocone (red curve) by more than two orders of magnitude with respect to the bare crystal (blue curve) at low intensity (Fig. 3a) and by around one order of magnitude at higher intensity (Fig. 3b). Fig. 3c and 3d report the intensity power law dependencies of H7 and H9, respectively. While H7 and H9 behave perturbatively with the laser intensity when generated in the bulk (with close to $I^q$ dependencies for harmonic order q), their yield scales non-perturbatively when enhanced in the nanocone.

To investigate the sustainability of the harmonic nano-source, we have followed the temporal evolution of the harmonic signal which exhibits unexpected dynamics. For intensities below one TW/cm$^2$ where resonant metallic nano-structures usually melt in few seconds[22,23], the harmonic signal emitted from the nanocone was enhanced during hours as shown in Figs. 4a and b for H7 and H9, respectively. After exposure of a pristine nanocone to the laser, the harmonic signal first increases during a time span of up to half an hour and then decreases and tends towards a constant value. The temporal evolution of the harmonic emission appears to be related to a structural modification of the nanocone, similarly to a study in bow-tie antenna recently reported[26]. Two deep subwavelength cracks appeared close to the center of the nanocone after half an hour of exposure (SEM-image in Fig. 4c)[27]. Due to the local field enhancement self-organized reshaping occurs which yields a higher field enhancement than in the initial pristine nanocone. This explains the initial increase of the harmonic signal. For longer exposure the reshaping continues and the top surface of the nanocone becomes porous (SEM-image in Fig. 4d), which leads to a decrease of the harmonic signal. We found that the damages occur at the position of highest field enhancement (Figs. 1b, c), i.e. in the center of the output surface of the nanocone, and that the linear cracks are perpendicular to the laser polarization. Those cracks are associated to emission of ions and electrons. We have further confirmed the field localization at the top of the nanoprobe by measuring the spatial distribution of hot electrons emitted during the strong field interaction (see Fig. 4e). The measurement is realized using photoemission electron microscopy (PEEM, see methods) with a spatial resolution of about 35 nm[28]. The central bright spot corresponds to the non-linearly photo-emitted electrons with an energy ranging from 6 to 12 eV and a source size of 590 nm in diameter. Note that this illustrates another potential of our semiconductor waveguide to create nanoscale source of hot electrons.

The strong localization of the electron emission reflects the nanoscale field enhancement inside the structure and is correlated to the harmonic source size. From the simulation shown in Fig. 1c, the transverse mode of the fundamental beam at the top of the nanocone has a size of about 900 nm. The diameter of H7 emission from the nanocone was measured to be 800 nm in diameter (see inset in Fig. 2). This opens perspectives in the realization of UV/EUV nano-probes for spectroscopy or imaging applications. As an example, HHG can be used to image nanoscale objects in a femtosecond snapshot[29]. Here, we selected H5 emitted from the nanocone by using a narrow bandwidth filter to investigate the feasibility of coherent diffractive imaging (CDI). A generic layout is displayed in Fig. 5. The beam is focused onto a sample consisting of a nanoscale pattern realized on a thin opaque membrane (75 nm $Si_3N_4$ + 150 nm Au) by focused ion beam milling. The coherence of the harmonic source allows to measure a well contrasted diffraction pattern in the far field. The incoherent part of the emission due to fluorescence creates a constant background that does not contribute to the diffraction pattern. The image of the sample was reconstructed using a phase retrieval hybrid input-output algorithm which acts as a virtual lens (see methods). The CDI reconstruction reproduces well the cross-shape of the sample with a spatial resolution of 1.6 µm. This proof-of-principle can be extended to the sub-100 nanometer range by using higher band gap material nanocone to push the HHG spectrum to shorter wavelengths[2]. Besides illustrating the coherence of the harmonics generated in the semiconductor nanocone, we propose the realization of a very compact ultrafast nanoscale microscope. Other techniques such as ptychography could allow generalizing the method to extended objects[30].

## Discussion

In conclusion, we have shown that high-harmonic generation can be confined in a single semiconductor 3D nano-structure. We have shown the severe competition between the enhancement of incoherent fluorescence versus coherent harmonics. We report that the fluorescence dominates harmonics around the zinc oxide band gap. At energies well above the band gap, we demonstrated a clear enhancement of harmonics by at least two orders of magnitude. In contrast to metallic nano-structures that have much less enhancement volume and get rapidly damaged, semiconductor nano-structures are more sustainable, with hours long lifetimes. Such nanoscale enhancement can replace external laser amplifiers using few nano-joules femtosecond laser systems to boost non-linear processes to the strong field regime. Single nano-emitters generate coherent UV/XUV radiation which offers a wide field of applications as, for example, ultrafast nanoscale imaging. Our nano-structure can offer also an appealing source of hot electrons of nanometer scale size, the localization of which is directly controlled by the confinement of the electric field. Finally, our demonstrations pave the way towards a stable source of novel all-solid-state attosecond optoelectronic devices operating at petahertz frequencies. Additionally, the control of the laser phase gradient should also favor tailoring the spatial and spectral phases of the harmonic emission allowing controlled emission of attosecond pulses[31].

## Methods

### Laser:

The seed mode-locked oscillator delivers a train of 1.3 ps hyperbolic secant squared (sech$^2$) shaped pulses centered at 1910 nm with a repetition rate of 19 MHz. The average power is 4 mW, corresponding to E = 0.2 nJ. The pulses are further amplified in a cladding-pumped amplifier based on LMA TDF to tens of nano-joules. Multisolitonic effects lead to the pulse fission followed by ejection of high-energy (8.7 nJ) pulses that are frequency-shifted to 2100 nm. The spectrum has a central wavelength of 2103 nm and was measured with an optical spectrum analyzer (AQ6375,

Yokogawa, Japan). The average power of the pulse train is 162 mW with a linear polarization state. The pulse duration was measured via autocorrelation to be 85 fs.

**Experimental setup:**

The laser is focused using a 25 mm focal length off-axis parabola either in bulk or nano-structured ZnO. Harmonic spatial profiles can be imaged in the near-field (crystal sample exit) by a lens (f = 60 mm) and a CCD-camera (Photonlines, PCO.Ultraviolet). Harmonic spectra can be detected from the near infrared down to the deep UV (Ocean Optics QE Pro, cooled). For the imaging experiment, H5 is selected with a FB420-10 transmission filter from Thorlabs and focused by a f = 50 mm lens onto the PCO.Ultraviolet CCD camera.

**Sample:**

The surface of a ZnO [0001] crystal is patterned with isolated truncated waveguide structures via focused ion beam (FIB) milling. The nanocone has a height of 6.1 µm, a bottom diameter of 4.1 µm and a top diameter of 2.6 µm. It is illuminated from the bottom with the 2.1 µm laser beam.

**Simulations:**

Finite difference time domain (FDTD) calculations using LUMERICAL Solutions have been performed to optimize the local enhancement of the electromagnetic field in the nanocone at a wavelength of 2.1 μm. The material of both the nanocone and its substrate is ZnO. Its optical parameters were imported from Bond et al., 1965, to LUMERICAL and used throughout the simulations. The surrounding medium is air. The structure is illuminated by a Gaussian beam with ω = 5.0 µm wave at 2.1 µm wavelength from below the substrate. We used perfectly matching layers at the boundaries. A finite mesh of 40 nm was used on x, y and z. The calculation time is about 3 hours on a bi-processors Intel Haswell 10C E5-2650V3 (10 cores, 20 threads, max. frequency 3 GHz, Bus speed 9.6 GT/s QPI, 768 GB registered SDRAM (DDR4 2133, 68GB/s bandpass)).

**Phase retrieval:**

The diffraction pattern was inverted using a hybrid input output algorithm. The reconstruction image is the coherent average over 50 independent runs of the algorithm, of around 500 iterations each. From the phase retrieval transfer function criterion, the spatial resolution is estimated at 1.6 µm.

**PEEM:**

Photoemission electron microscopy measurements are carried out on a LEEM/PEEM III instrument (Elmitec GmbH) operating under ultra-high vacuum conditions. In PEEM imaging mode the spatial resolution routinely achieved is 20 nm. The laser source is a mode-locked Ti:Al$_2$O$_3$ oscillator (Chameleon Ultra II, Coherent Inc., repetition rate 80 MHz, pulse width 130 fs, wavelength range 680 nm – 1000 nm). In addition to PEEM imaging, the instrument can be operated in low energy electron microscopy (LEEM) mode where backscattered electrons are used to create an image reflecting the topography of the sample.

## Acknowledgments:

We acknowledge Franck Fortuna and Laurent Delbecq for access and support to the nano-focused ion beam at the CSNSM laboratory (IN2P3, Paris Saclay University).

## Funding:

We acknowledge support from the PETACom FET Open H2020 support, from the French ministry of research through the ANR grants 2014 "IPEX", 2016 "HELLIX", 2017 "PACHA", the DGA RAPID grant "SWIM" and from the C'NANO research program through the NanoscopiX grant, and the LABEX "PALM" (ANR-10-LABX-0039-PALM) through the grants "Plasmon-X", "STAMPS" and "HILAC". We acknowledge the financial support from the French ASTRE program through the "NanoLight" grant. Financial support by the Deutsche Forschungsgemeinschaft, grant KO 3798/4-11 and from Lower Saxony through "Quanten- und Nanometrologie" (QUANOMET), project NanoPhotonik are acknowledged. J.B. acknowledges financial support from the Spanish Ministry of Economy and Competitiveness (MINECO), through the "Severo Ochoa" Programme for Centres of Excellence in R&D (SEV-2015-0522) and the Fundació Cellex Barcelona.

## Author contributions:

D.F., S.K., D.G., R.N., W.B., and H.M. carried out the experiment. The samples were produced by W.B.. The laser was provided by N.D, S.F., J.-T. G., L.L., and D.Gap. Simulations were performed by D.F., S.K., R.N. and D.G.. The PEEM measurements were made by L.D.. Data analysis was performed by D.F., S.K., D.G. R.N., M. Kh., W.B., J.B., L.S., M.K. and H.M. H.M. proposed the physical concept. All authors discussed the results and contributed to the writing of the manuscript.

## Competing interests:

The authors declare no competing financial and non-financial interests.

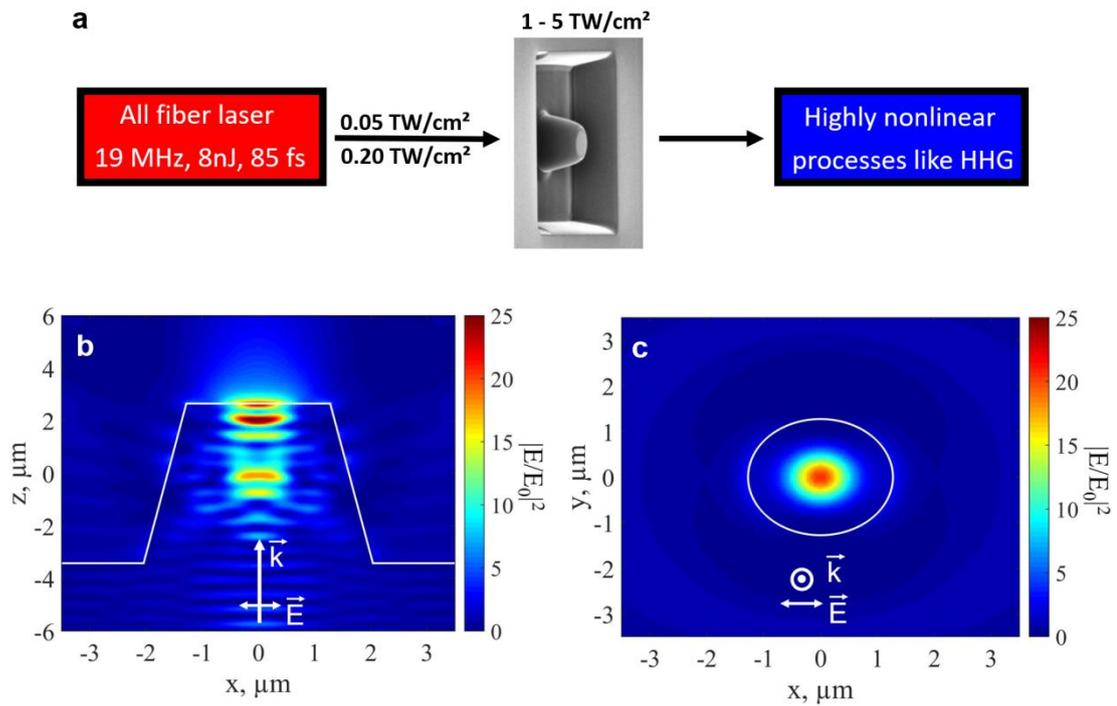

**Fig. 1. Field enhancement in the semiconductor nanowaveguide. a** Schematics of the field enhancement. The local field enhancement increases moderate pump intensities of 0.05 – 0.2 TW/cm$^2$ provided by the laser to up to 5 TW/cm$^2$ and allows for HHG. **b** Simulation of the intensity distribution in the nanocone (plane spanned by k- and E-vectors). Significant intensity enhancement occurs on the tip of the nanocone by a factor larger than 20. The structure is patterned on a bulk ZnO crystal with the dimensions: base size of 4.1 µm in diameter, height of 6.1 µm and top of 2.6 µm. The simulation has been done with the experimental parameters. **c** Intensity enhancement at the exit of the nanocone (plane perpendicular to the propagation direction) 100 nm below the top surface of the nanocone.

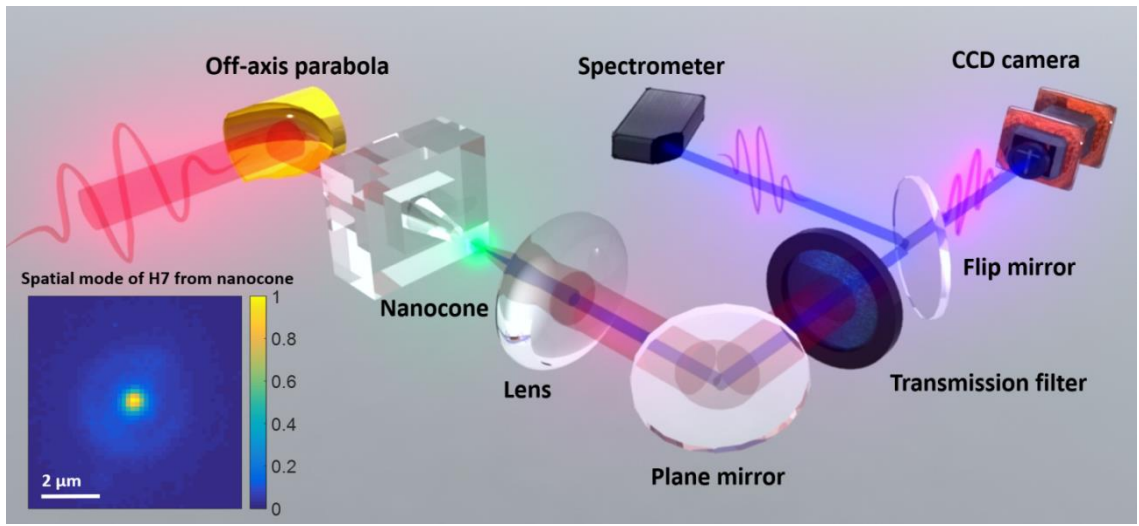

**Fig. 2. Experimental setup.** The laser used in this experiment is an all fiber laser with 2.1 µm operating wavelength, 85 fs pulse duration, 19 MHz repetition rate and 8.7 nJ maximum pulse energy. The laser is focused with an off-axis parabola (f = 25 mm) into the sample to a spot size of 5.9 µm (FWHM) at normal incidence. The intensity range used in the experiment is 0.05 – 0.2 TW/cm$^2$. The light exiting the nanocone is collected by a lens towards a spectrometer or a CCD camera. Different transmission filters can be used to select a specific harmonic order. The spatial profile of H7 emitted from the nanocone is shown as an inset.

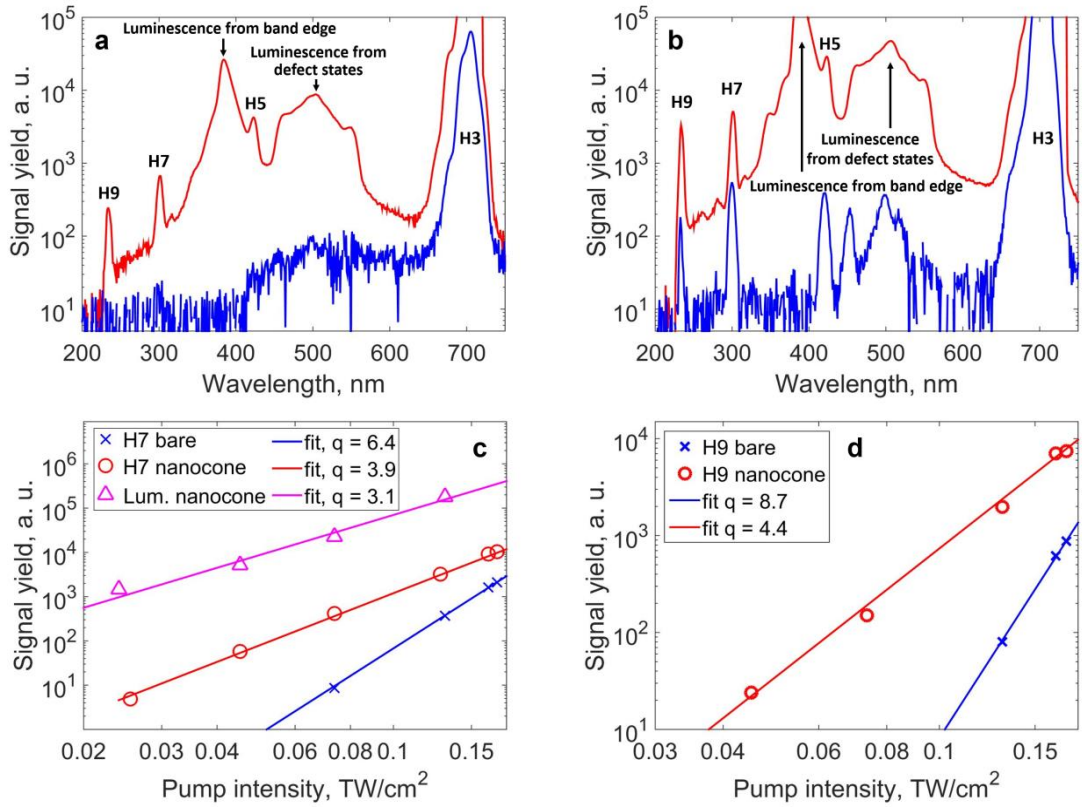

**Fig. 3. Intensity-dependent signal yield. a** and **b** Spectra emitted from the ZnO nanocone (red curve) and from the bare crystal (blue curve) measured at pump intensities of 0.07 TW/cm$^2$ and 0.17 TW/cm$^2$, respectively. The harmonics H3, H5, H7 and H9 are highlighted as well as the luminescence from the band edge at 385 nm and from defect states. **c** and **d** Intensity dependence and power law of the harmonic yield for H7 and H9 from the bare crystal (blue crosses) and the nanocone (red circles), respectively. The power law for the luminescence from the band edge generated in the nanocone is shown in **c** as magenta triangles.

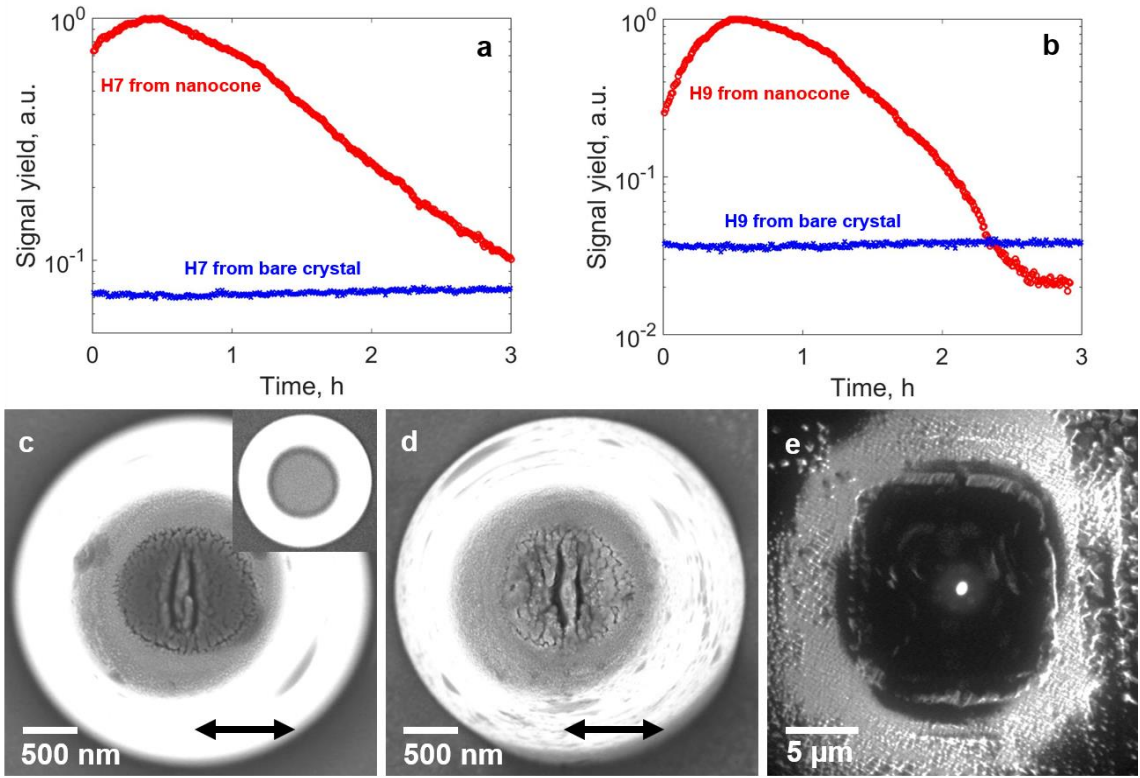

**Fig. 4. Temporal evolution of the harmonic emission.** Temporal evolution of **a** H7 and **b** H9 from a nanocone at an intensity of 0.17 TW/cm$^2$ (red curve). The signal is observed over a time of 3 hours. SEM-images of the top of a nanocone after an exposure time of half an hour and 3 hours, respectively at an intensity of 0.17 TW/cm$^2$ are shown in **c** and **d**. The polarization direction is indicated by a black arrow. The initial nanocone, before irradiation, is shown as an inset in **c**. **e** Image of hot electrons emitted by the nanocone taken by photo-emission electron microscopy (PEEM). The nanocone is back illuminated by a 775 nm pulsed laser beam under an angle of 45°. The image is a superposition of two simultaneously acquired images, one recorded in low energy electron microscopy imaging mode (LEEM, topographic signature) and one in PEEM imaging mode (photoelectron signature).

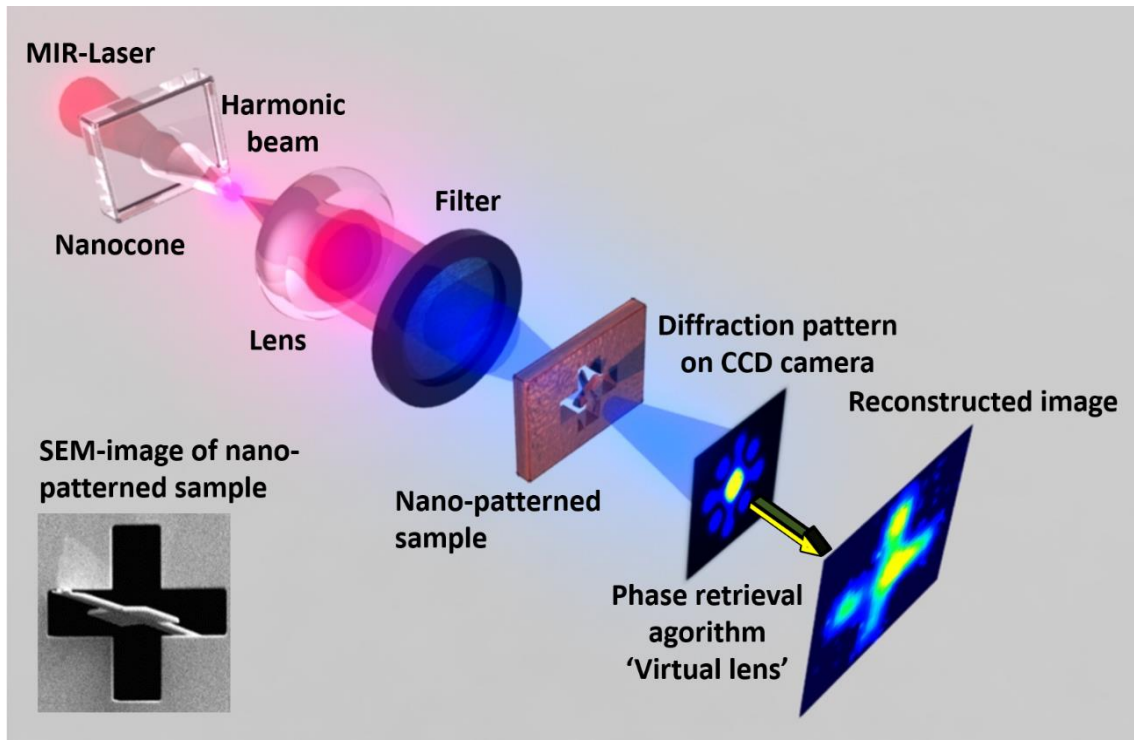

**Fig. 5. Generic layout of the coherent diffractive imaging setup.** Few nano-joules of our 2.1 μm femtosecond laser are injected in the nanocone. The emission from the nanocone was filtered in the far field to select the 5th harmonic and then focused on a nano-patterned test sample consisting in a partially occulted cross. The diffraction pattern is collected in the far field using a CCD detector. A phase retrieval algorithm is then applied to reconstruct the image of the sample.